\documentclass[preprint,authoryear,times,12pt]{elsarticle}

\usepackage{amssymb,amsmath}

\usepackage{lineno}

\usepackage{graphicx}
\usepackage{xspace}
\usepackage{bm}
\usepackage{longtable}
\usepackage{hyphenat}

\newcommand{\mtl}{$\bar{t}$\xspace}
\newcommand{\tf}{\textbf{TransmissionFramework}\xspace}
\newcommand{\ct}{cultural transmission\xspace}

\newcommand{\timeav}{time averaging\xspace}
\newcommand{\timeavd}{time averaged\xspace}

\newcommand{\Timeav}{Time averaging\xspace}

\usepackage[bookmarks=true,linkcolor=blue,hyperfootnotes=false,breaklinks=true,citecolor=blue,colorlinks=true]{hyperref}
\hypersetup{pdftitle={Unbiased Cultural Transmission in Time Averaged Archaeological Assemblages},pdfauthor={Mark E. Madsen}}

\journal{Journal of Anthropological Archaeology}

\begin{document}

\begin{frontmatter}

\title{Neutral Cultural Transmission in Time Averaged Archaeological Assemblages}

\author{Mark E. Madsen}

\address{Department of Anthropology, Box 353100, University of Washington, Seattle WA, 98195 USA}
\ead{mark@madsenlab.org}
\ead[url]{http://madsenlab.org}

\begin{abstract}
Neutral models are foundational in the archaeological study of cultural transmission.  Applications have assumed that archaeological data represent synchronic samples, despite the accretional nature of the archaeological record.  Using numerical simulations, I document the circumstances under which time-averaging alters the distribution of model predictions.  Richness is inflated in long-duration assemblages, and evenness is ``flattened'' compared to unaveraged samples.  Tests of neutrality, employed to differentiate between biased and unbiased models, suffer serious problems with Type I error under time-averaging.  Estimation of population-level innovation rates, which feature in many archaeological applications, are biased even without \timeav, but have sharply increased bias given longer assemblage durations.  Finally, the time scale over which time averaging alters predictions is determined by the mean trait lifetime, providing a way to evaluate the impact of these effects upon archaeological samples.  
\end{abstract}

\begin{keyword}
cultural transmission \sep Wright-Fisher model \sep time averaging \sep neutral theory
\MSC[2010]{91D99}
\end{keyword}

\end{frontmatter}


\section{Introduction}
 
The evolutionary study of culture today crosses many disciplines and employs a variety of experimental and observational methods to study its subject matter.  What makes the archaeological record unique as a source of data concerning the evolution of culture is time depth, creating the possibility of studying both the unique histories of human groups and the evolutionary processes that shape those histories.  Archaeology is not unique in studying temporal data on human activity, but like our colleagues in paleobiology, we study an empirical record that is unlike the time-series data available to disciplines such as economics or epidemiology \citep[e.g.,][]{arrow2009some,keeling2005implications,keeling2007modeling,kendall1953analysis,rothman2008modern}. The archaeological record is not a sample of measurements from individual moments in time stacked together into a sequence.  Instead, archaeological deposits are almost always accretional palimpsests, representing cumulative artifact discard over durations of varying length \citep{bailey2007time,bailey1981concepts,binford1981behavioral,8981,stein1987deposits}.  Thus, when archaeologists count the richness of faunal taxa in an assemblage, or measure the relative frequencies of ceramic types, the data obtained summarize the bulk properties of artifact discard and deposition over significant spans of time, often with nonconstant rates of accumulation.\footnote{As well as the action of various post-depositional and taphonomic processes, of course.}  We refer to assemblages which are accretional in this manner as ``\timeavd.''  

A growing number of studies apply \ct models to artifact assemblages by comparing the predictions such models make for the richness, diversity, or frequency distribution of cultural traits, to counts or frequencies of artifact classes \citep[e.g.,][]{Bentley2003,bettinger1999point,Eerkens2007,8994,Lipo2000,perreault2010mobility,premo2011spatial,scholnick2010apprenticeship,shennan2001ceramic,shennan2011descent,steele2010ceramic}.  The question is, are model predictions comparable to archaeological measurements?  Given the \timeavd structure of most archaeological deposits, I suspect the answer is no.  Transmission models developed outside archaeology are typically constructed to make predictions concerning variables observed at a point in time.  To date, almost none of the archaeological literature employing \ct models has taken this ``\timeav'' effect into account and modified the way predictions are made to match the nature of the phenomena we measure \citep[cf.][]{bentley2004random}.  Evaluating the effects of temporal aggregation upon the predictions made by \ct models is the first step in understanding how to rewrite and adapt transmission models to understand their dynamics given \timeavd observations.  

In his dissertation, \citet{Neiman1990} considered a potential source of \timeav effects in diachronic assemblages:  variation in discard rates across traits.   With respect to this particular effect within accretional deposits, Neiman's results suggested that the predictions made by a neutral model of \ct were directly applicable to the relative frequencies of traits as we would measure them in a \timeavd assemblage.  Nevertheless, there is good reason to consider the effects of aggregation directly, outside of variation in discard rates.   
Paleobiologists, for example, have documented systematic differences between living and fossil assemblages, including increased species richness, reduced spatiotemporal variance in taxonomic composition, and flatter species abundance curves in \timeavd assemblages \citep{olszewski2011remembrance,tomasovych2010effects,tomasovych2010predicting}.  \citet{lyman2003influence} extended these results to zooarchaeology, noting that \timeav can be a significant problem when the process one is applying or studying occurs over a shorter time scale than the empirical record available to study its properties \citep[see also][]{grayson1998}.  This relation between time scales is applicable to \ct modeling as well.  

Archaeologists now employ a variety of \ct models, which differ in the kind of variation and traits they describe and the copying rules and evolutionary processes they incorporate.  Discrete models describe individual variants or traits by their count or frequency in a population and are foundational for the study of stylistic variation in many artifact categories (e.g., pottery).  The simplest discrete  model is random copying in a well-mixed population with innovation, representing neutral variation with the stochastic effects of drift.  We frequently construct more complex models of transmission bias by adding additional terms or frequency-dependent copying rates to the basic unbiased copying model \citep{cavalli1973cultural,cavalli1973models,CF1981,BR1985}.  Thus, an understanding of the effects of \timeav upon neutral transmission will be informative about many (if not all) of the discrete transmission models in use by archaeologists today, and forms the focus of the present study.  

I report the results of numerical simulations designed to observe neutral transmission using variables employed in the archaeological literature, aggregated over time at a variety of intervals designed to mimic a wide range of ``assemblage durations.''  In Section \ref{sec:concept-review} I describe the relationship between neutrality, unbiased copying, and the separate but related concept of ``drift,'' followed by a review of the quantitative properties of the well-mixed neutral Wright-Fisher infinite-alleles model in Section \ref{sec:wf-model}.  Section \ref{sec:methods} outlines the simulation model employed to study \timeav in this paper, including model verification and testing, and the algorithm used to effect temporal aggregation within the simulations.  Section \ref{sec:results} presents the results of simulating unbiased \ct for a variety of innovation rates and assemblage durations, and Section \ref{sec:conclusions} summarizes the effects seen and points to next steps in reformulating our \ct models for archaeological contexts.

\section{Conceptual Structure of Neutral Cultural Transmission}
\label{sec:concept-review}

In his classic article “Style and Function: A Fundamental Dichotomy,” Dunnell \citeyearpar{8961} proposed that many aspects of an artifact would play little or no role in its engineering performance, and thus have no impact on the fitness of individuals employing it. In other words, some attributes of artifacts are neutral with respect to selection.  This has been widely misinterpreted as a claim that the artifacts themselves are neutral or have no fitness value, which is not the case. Dunnell was saying that if one describes an artifact solely using attributes which have equal cost or performance, the resulting classes meet the definition of neutral variation.  

Fraser Neiman \citeyearpar{Neiman1990} first connected Dunnell's identification of style as selectively neutral variation, to population genetic models designed to describe genetic drift.  His dissertation considers a wide range of \ct models, especially those described by \citet{cavalli1973cultural,cavalli1973models,CF1981} and \citet{BR1985}.  Neiman employed simulation to calculate the consequences of both individual processes as well as processes combined with various archaeological factors such as variable rates of artifact discard.  In this work, Neiman pioneered virtually every technique used by archaeologists today to model and study cultural transmission.  The discipline as a whole was introduced to this work in his now classic 1995 article \citep{Neiman1995}, in which the dynamics of Woodland ceramic variation were explicitly modeled as a random copying process.

Despite the fact that there are multiple ways that neutrality can arise as a population level effect, there is a tendency today to equate neutrality with ``drift'' in the archaeological literature on \ct.   For example, \citet[][p.1443]{bentley2004random} offer a fairly typical description of unbiased \ct as  ``random genetic drift, which describes how the diversity of variants evolve when the dominant process is one of random copying.''  In fact, drift and the copying rules that create population-level trait distributions are different and independent aspects of a transmission system.   Before we turn to the details of a formal model for unbiased, neutral transmission, it is worth reviewing the conceptual elements that make up such models.

Drift is a feature of any stochastic transmission model in a finite population, regardless of whether selection or bias is also present in the model.   Sewall Wright gave the name ``genetic drift'' to the random fluctuations in gene frequency that occurred because some individuals might be the source of many genes in the next generation, and others none at all.  Translated into a cultural model, drift occurs when some individuals, by random chance, are imitated or copied and others are not.  In an infinite population, by contrast, the variants held by individuals would be sampled at their exact frequencies in the population, and thus there would be no stochastic ``wiggle'' in trait frequencies.  This is reflected in population genetics by the famous ``Hardy-Weinberg'' equilibrium, where in the absence of selection or other forces, gene frequencies stay the same from generation to generation.    This means that we can easily have \emph{neutrality without drift}, in an infinite population.  In a large but still finite population, we can expect drift to have very tiny, potentially even unmeasurable effects upon the trajectory of trait frequencies.   

Drift, moreover, occurs in combination with a variety of inheritance rules, mutation models, and in combination with natural selection.  In small populations, we can expect drift to be a factor when examining the engineering properties of ceramics and the relative fitness of firing technologies, or the fitness of foraging strategies.  Whenever such traits are learned and passed on within small, finite populations, the stochastic aspect of who learns from whom will create fluctuations in variant frequencies that have nothing to do with the performance or survival value of traits, or the prestige of those we choose to learn from or imitate.  In other words, we can have \emph{drift without neutrality}.  In small enough populations or during bottlenecks, even adaptive technologies and knowledge can be lost to drift \citep{8921,henrich2006understanding}.  We should always be on the lookout for the effects of drift, especially as population sizes get smaller as we go back in time.  Drift is not a model of human social learning; it is a consequence of finite populations, injecting stochastic noise into the dynamics of a system that affects our ability to cleanly fit models and test hypotheses.  

Neutrality, by contrast, is a population level phenomenon, arising when there is no net force systematically favoring certain variants over others for a particular dimension of variation.  Most commonly, of course, we mean that there is no natural selection that favors some alleles over others, but from a mathematical perspective, the transmission bias rules of  \citet{CF1981} and \citet{BR1985} are equivalent to selection models.\footnote{In this paper I leave aside the relationship between ``natural'' and ``cultural'' selection, and transmission biases, since such issues are largely philosophical and theoretical and do not affect the nature of the models we employ for quantitative analysis of cultural variation.}   The simplest way for neutrality to arise is for individual social learning to be ``unbiased.''  Unbiased transmission models always yield population-level neutrality for the traits being passed, because the probability of imitating any specific trait is simply proportional to its frequency among individuals in the population.  The Wright-Fisher model is one of the earliest stochastic models in population genetics \citep{provine1989sewall,provine2001origins,wright1931evolution}, and was originally created to describe the process of genetic drift and its effects in combination with other evolutionary processes.  Following Kimura's theory of neutral alleles, Wright-Fisher is also used to describe the evolution of populations in which variants are selectively neutral.  Elaborations of the basic Wright-Fisher model add mutation, selection, loci with multiple alleles, and multiple loci with interactions between loci \citep[see esp.][]{crow1970introduction,Ewens2004}.\footnote{And, the Moran family of models mirrors the Wright-Fisher models, with overlapping generations, by representing dynamics as continuous-time stochastic processes.  Moran models are likely the best framework for modeling \ct when the exact temporal dynamics matters.  In this paper I follow archaeological convention by employing the more familiar Wright-Fisher discrete generation framework.}
 
But unbiased copying is not the only source of neutrality among variants, and it is important to keep this in mind when selecting models to test as explanations for archaeological phenomena.   In any realistic human population, there will be heterogeneity in social learning rules, with individuals using different rules for different traits, or kinds of traits, and perhaps having individual propensities for conformism (all other things being equal) or pro-novelty bias \citep{Mesoudi2009}.  A population which is heterogeneous for such rules may display the characteristic frequency distributions of conformity or pro-novelty biased if we are able to observe small numbers of transmission events or individual transmission chains, while simultaneously cancelling each other out at the level of the population.  In other words, heterogeneity is a major source of equifinality between different models of social learning, when observed through population-level trait frequencies.   No archaeological applications of \ct models today have employed heterogeneous models, probably because the theory behind such models is not well-studied.  But this is clearly a frontier for future research since homogenous models poorly reflect what occurs in real human populations.  

Returning to unbiased models of transmission, we face a further choice in selecting a specific model to employ or study.  In addition to the copying rules, we must specify an innovation rule.  Such a rule answers questions like:  how do new variants enter the population, can variants be invented multiple times independently, and is there a constrained range of variation for a particular dimension of an artifact?  For example, painted design elements on a ceramic pot offer a ``design space'' of possibilities that is potentially unbounded, even if only a tiny fraction of possible designs occur in any archaeological context.  Such attributes are best modeled by the ``infinite allleles'' innovation model.  In contrast, stylistic aspects of lithic tools may be sharply constrained by the technology and materials themselves, and may be best modeled by innovation among a small set of variants, with the material constraints causing frequent ``reinvention'' of the same shapes over and over.   Such attributes are best modeled by constraining the design space, and employing a finite or ``k-alleles'' version of the unbiased model.   Since Neiman's pioneering work, most archaeological applications of neutral models have employed the ``infinite alleles'' variant of the Wright Fisher model  (WF-IA)\citep{kimura1964number}.  Therefore, in the remainder of this paper, I focus on the unbounded model of neutral evolution with innovation, since it is relevant to a large number of archaeological contexts and artifact categories, but the reader should be aware that the models with a constrained number of variants may be hugely important in specific archaeological contexts, and are underexplored in the archaeological literature.  

\section{Unbiased Transmission:  The Wright-Fisher Infinite-Alleles Model}
\label{sec:wf-model}

WF-IA is a stochastic process that models unbiased transmission within a fixed-size population as multinomial sampling with replacement, with a mutation process that adds new variants to the population at a known rate.  After describing the model, I review the sampling theory of \citet{ewens1972sampling}, which gives the distribution of variants expected in small samples taken from the population as a whole.  The sampling theory, rather than the distribution of variants in the full population, is both well-understood, and most relevant to archaeologists, who are always sampling an empirical record of past artifact variation.

The well-mixed neutral Wright-Fisher infinite-alleles model \citep{kimura1964number} considers a single dimension of variation (``locus'') at which an unlimited number of variants (``alleles'') can occur, in a population of $N$ individuals.\footnote{Conventionally, the model treats a diploid population, in which N individuals each have two chromosomes and thus there are always 2N genes tracked in the population.  The haploid version is more appropriate for modeling cultural phenomena, and thus formulas given in this paper may differ from those given by \citet{Ewens2004} and other sources by a factor of two.  For example, the key parameter $\theta$ is defined as $2N\mu$ rather than the common genetic definition $4N\mu$.}  The state of the population in any generation is given in several ways:  a vector representing the trait possessed by each individual (census), a vector giving the abundance of each trait in the population (occupation numbers), or by the number of traits represented in a population by a specific count (spectrum).  

In each generation, each of $N$ individuals selects an individual at random in the population (without respect to spatial or social structure, hence ``well-mixed''), and adopts the trait that individual possessed in the previous generation.\footnote{An individual can select themselves at random since sampling is with replacement, and this would be equivalent to ``keeping'' one's existing trait for that generation.}  Equivalently, a new set of $N$ individuals are formed by sampling the previous generation with replacement.  At rate $\mu$ for each individual, a new variant is added to the population instead of copying a random individual, leading to a population rate of innovations $\theta = 2N\mu$ \citep{Ewens2004}, with no ``back-mutation'' to existing traits.\footnote{It is important to note that $\theta$ is not a measure of the ``diversity'' of traits in the population, as it has been employed in several archaeological studies, but is instead a \emph{rate} parameter of the model.}  An important consequence of this innovation model is that each variant is eventually lost from the population given enough time, and replaced with new variants.  Thus, there is no strict stationary distribution for the Markov chain describing WF-IA, although there is a quasi-stationary equilibrium in which the population displays a characteristic number of variants, with a stable frequency distribution governed by the value of $\theta$ \citep{Ewens2004,watterson1976stationary}.   

Beginning with a now-classic paper \citet{ewens1972sampling} constructed a sampling theory for the neutral WF-IA model, allowing the calculation of expected moments and frequency distributions for small samples (compared to overall population size) \citep[see ][for a complete summary of results on the sampling theory]{Ewens2004}.  In what follows, we assume that a neutral WF-IA process is running within a population of size $N$.  At some moment in time after the population has reached its quasi-stationary equilibrium, we take a sample of $n$ individuals, where the sample is small compared to the population size ($n \ll N$).  We then identify the variants held by each individual.  The total number of variants seen in the sample will be denoted by $k$, or $k_{\text{obs}}$ depending upon context.     

Given such a sample, \citet{ewens1972sampling} found that the joint distribution of the variant spectrum ($a_i$ represents the number of variants represented $i$ times in a sample), given the population innovation rate ($\theta$), is given by the following formula (now known as the Ewens Sampling Distribution):

\begin{equation}
\label{eq:esd}
\mathbb{P}_{\theta,n}(a_i, \ldots, a_n) = \frac{n!}{\theta^{(n)}} \prod^n_{j=1} \frac{(\theta/j)^{a_j}}{a_j!}
\end{equation}

where $\theta^{(n)}$ is the Pochhammer symbol or ``rising factorial'' $\theta(\theta+1)(\theta + 2)\cdots(\theta + n - 1)$.  In most empirical cases, we cannot measure (or do not set through experiment) the value of $\theta$, so a more useful relation is the distribution of individuals across variants (i.e., the occupation numbers), conditional upon the number of variants $k_{\text{obs}}$ observed in a sample of size $n$:

\begin{equation}
\label{eq:conditional-esd}
\mathbb{P}(n_1, n_2, \ldots, n_k | k_{obs}) = \frac{n!}{|S^k_n| k! n_1 n_2 \cdots n_k}
\end{equation}

where $|S^k_n|$ denote the \emph{Stirling numbers of the first kind}, which give the number of permutations of $n$ elements into $k$ non-empty subsets \citep{abramowitz1965}.  The latter serves here as the normalization factor, giving us a proper probability distribution.   

From Ewens's sampling theory, and in particular Equation \ref{eq:conditional-esd}, a number of useful measures can be derived, relevant to archaeological applications.  In this study, I focus upon the most commonly used:  statistical tests of neutrality, estimation of innovation rates ($\theta$), and the evenness with which variants are represented in the population (as revealed by several diversity measures).  

\subsection{Statistical Tests for Neutrality}
\label{sec:neutrality-test}

Because Equation \eqref{eq:conditional-esd} requires no unobservable parameters, it serves as the basis for goodness-of-fit tests between empirical samples and the neutral WF-IA.  The two most important such tests are the Ewens-Watterson test using the sample homozygosity and Slatkin's ``exact'' test \citep{durrett2008,Ewens2004,slatkin1994exact,slatkin1996correction,slatkin1994exact,slatkin1996correction}.\footnote{There are several other important tests of neutrality when dealing with DNA sequence data, including Tajima's D, the HKA test, and the McDonald-Kreitman test \citep{durrett2008}.  Because their assumptions are highly specific to the structure of sequence data, I omit consideration of them here.}  Both have been adopted for use by archaeologists, beginning with \citet{Neiman1995} and \citet{Lipo2001b}, who described Watterson's work in detail, and more recently, applications of Slatkin's exact test by \citet{steele2010ceramic} and \citet{premo2011spatial}. 

The Slatkin test makes no assumptions concerning the process underlying an alternative hypothesis to neutrality, whereas the Ewens-Watterson test examines the observed heterozygosity at a locus versus the expected heterozygosity predicted by Ewens sampling theory.  Slatkin's test does not employ the concept of heterozygosity, and relies only upon the ``shape'' of the Ewens Samping Distribution given a specific innovation rate.  As a result, archaeologists should prefer Slatkin's test for examining the fit of a synchronic sample of variants to the null hypothesis of neutrality.  Slatkin's test is modeled upon the Fisher exact test for contingency tables.  Where the Fisher exact test determines the probability of an unordered configuration from the hypergeometric distribution, Slatkin's test determines the probability of a sample of traits (characterized by occupation numbers) with respect to Equation \ref{eq:conditional-esd}.  

There are two methods for determining how probable a given sample is, with respect to the ESD.  For relatively small $n$ and $k$, it is possible to enumerate all possible combinations ($\mathbf{C}$) of the $n$ individuals among $k$ variants.  Each configuration ($c_j \: \in \: \mathbf{C}$) then has a probability given Equation \ref{eq:conditional-esd}, as does the observed configuration ($c_{\text{obs}}$).   With larger sample sizes and values of $K_{\text{obs}}$, it becomes impractical or simply time consuming to enumerate all possible configurations and thus determine the likelihood of an observed sample.  In such cases, Monte Carlo sampling of configurations from the Ewens Sampling Distribution is used.  We then determine the total probability mass of all configurations (enumerated or sampled) whose probability are less than or equal to the observed configuration:

\begin{equation}
\label{eq:slatkin-pe}
\mathbb{P}_e = \sum_{c_j \in \lbrace \mathbf{C} \: : \:  P(c_j \: | \: k) \; \leq \; P(c_o \: | \: k)\rbrace} \mathbb{P}(c_j \: | \: k)
\end{equation}

$\mathbb{P}_e$ then represents the Fisherian p-value of the sample with respect to the Ewens Sampling Formula, and thus can be interpreted as a test of the hypothesis that the sample was drawn from a neutral dimension of variation which followed the WF-IA copying model.  The $\mathbb{P}_e$ value for a given sample gives the tail probability of its occurrence given the ESD.  Thus, if we take a sample of size 100 in a population with innovation rate $\theta =  0.1$, and identify two variants with counts 51 and 49, we might not be surprised to see a $\mathbb{P}_e$ value of 0.01181, indicating that such a sample is highly unusual for a WF-IA process.  On the other hand, in the same sample of size 100, if we identify four variants, with counts 55, 38, 6, and 1, this seems a much more typical result of an unbiased copying process.  Indeed, the $\mathbb{P}_e$ value of 0.48544 confirms that we should expect to see such samples quite often.

\subsection{Estimation of Innovation Rates}
\label{sec:theta-estimation-theory}

The behavior of the WF-IA neutral model is governed by the innovation rate ($\theta$).  Recall that $\theta = 2 N \mu$, and thus represents the population-level rate at which new variants enter the population.  In general, for low values of the innovation rate ($\theta < 1.0$), the process is ``drift-dominated,'' and one or a small number of variants dominate the population.  At innovation rates above 1.0, which implies that every single ``generation'' incorporates one or more new variants, the process is ``mutation-dominated,'' and more variants are maintained at intermediate frequencies in the population.  

Thus, estimation of the innovation rate from empirical data is of great interest when investigating empirical cases.   If we measure the number of variants ($K_n$) in a sample of artifacts of size $n$, the sampling theory  gives the following probability distribution \citep[Eq. 3.84]{Ewens2004}:

\begin{equation} 
\label{eq:full-distro-kn}
	\mathbb{P}_{\theta}(K_n = k) = \frac{|S^k_n| \theta^k}{\theta^{(n)}}
\end{equation}

This is a somewhat inconvenient distribution to work with directly, since calculating the Stirling numbers and rising factorials is both analytically difficult and computationally expensive, but the expected value of $K_n$ has a simple form:

\begin{equation} 
\label{eq:expected-kn}
	\mathbb{E}(K_n) = \frac{\theta}{\theta} + \frac{\theta}{\theta + 1} + \frac{\theta}{\theta + 2} + \cdots + \frac{\theta}{\theta + n - 1}
\end{equation}

$K_n$ is the sufficient statistic for $\theta$, containing all of the information required to calculate the maximum likelihood estimate of the innovation rate ($\hat{\theta}$) from an empirical sample.  This is done numerically by finding the value of $\theta$ that maximizes the likelihood function of Equation \ref{eq:full-distro-kn}, or equivalently, finding the value of $\theta$ for which the expected value of $K_n$ given Equation \ref{eq:expected-kn} is equal to the observed number of variants in a sample (since the full distribution may not have a closed-form likelihood function).  In the archaeological literature, \citet{Neiman1995} introduced this estimator of $\theta$ and called it $t_e$.  With larger samples, \citet{watterson1975number} showed that $k\;/\;\log n$ is a good approximation for the MLE estimator \citep{durrett2008}.  

Despite the fact that this estimator (and its approximations) are the best that can be achieved from samples, \citet{ewens1972sampling} showed that all such estimates of $\theta$ are biased.  Simulations demonstrate, furthermore, that $\hat{\theta}$ (or $t_e$) is an overestimate of the actual value, and that the amount of bias increases with $\theta$ itself \citep{ewens1974some}.   In addition, the variance of the estimator is quite large, and decreases very slowly with increased sample size \citep{durrett2008}.  The situation is quite different using the ``infinite sites'' model of neutral evolution and DNA sequence data, where there are excellent and nearly unbiased estimators of theta.  

But with the WF-IA and no additional structure to ``traits'' or alleles, it is very difficult to estimate the innovation rate with any accuracy, or determine whether two samples come from populations with the same innovation rate, or different rates.  This fact calls into serious question the degree to which $t_e$ is useful in archaeological analysis, either for estimating innovation rates in past populations, or as a measure of richness or diversity across assemblages or samples.  These caveats apply to estimates of innovation rates and $t_e$ given synchronic samples; the effects of \timeav on theta estimation have not been previously documented, and are addressed in Section \ref{sec:theta-estimation-results}.

\subsection{Diversity Measures}

The amount of variation expected in a sample is an important quantity, given that we would clearly expect transmission models incorporating bias terms to differ from unbiased or neutral models \citep[e.g.][]{8977}.  Conformist transmission should result in smaller numbers of variants than expected under unbiased transmission, and of course anti-conformist, or ``pro-novelty'', biases should result in larger numbers of variants being maintained, on average.  But beyond helping us assess goodness-of-fit to an unbiased copying model, comparing the number of variants in a sample ($K_n$) either to a model, or between assemblages, is difficult without reliable estimates of the population-level innovation rate ($\theta$).  Since this is inherently difficult and inaccurate, we might ask instead what the evenness of variants is across our samples, since both innovation rates and different models of \ct have clear implications for the diversity of traits we observe.  

In the archaeological literature on \ct, the most important evenness measure is $t_f$, which is a summed estimate of dispersion given trait frequencies \citet{Neiman1995}:  

\begin{equation}
\label{eq:tf-formula}
t_f = \frac{1}{\sum_{i=1}^k p_i^2} - 1
\end{equation}

To make this measure easier to compare across different innovation rates, it is convenient to normalize.   Wilcox's ``index of quantitative variation,'' does so, and varies between 0 (when all cases belong to a single category), and 1 (when all cases are evenly divided across categories) \citep{wilcox1973indices}:

\begin{equation}
\label{eq:iqv-formula}
\mathrm{IQV} = (\frac{k}{k-1}) (1 - \sum_{i=1}^k p_i^2 )
\end{equation}

Paleobiologists have found that fossil assemblages have considerably ``flatter'' species diversity curves compared to living communities, and I expect that \timeav will have the effect here of pushing $\mathrm{IQV}$ towards 1.0 compared to its value in unaveraged samples.  


\section{Methods}
\label{sec:methods}
In this research, I employ a ``forward-time'' approach to computational modeling of unbiased \ct, by contrast to most modeling in theoretical population genetics today, which employs the coalescent or ``backward-time'' approach \citep{kingman1977population,durrett2008,wakeley2008}.  In archaeological research, we are interested in the entire distribution of variants which transmitted through the population, samples of which may be deposited and become part of the archaeological record regardless of which variants ultimately leave descendants in later generations.  Forward-time approaches evolve a population in steps, applying rules for the generation of variation, copying between individuals, innovation, and sometimes population dynamics.\footnote{Forward-time approaches are not necessarily equivalent to ``agent-based models,'' but ABM techniques are useful in implementing forward-time models.}  Several well-tested forward-time population genetic frameworks exist, including a very flexible framework called \textbf{simuPOP} \citep{peng2012forward,peng2005simupop}.  

In this research, I employ a framework written by the author specifically for \ct simulations.  This project calls for integrating computation models of archaeological classification and seriation, which require code beyond that supplied by population genetics frameworks.  My simulation codebase is called \tf, and is available as open-source software.\footnote{\tf can be downloaded or the code examined at \url{http://github.com/mmadsen/TransmissionFramework}.}  \tf runs on any platform capable of supporting a Java 1.6+ runtime, with optional scripts requiring Ruby 1.9+.

\subsection{Model Verification}
\label{sec:verification}

Simulation modeling plays an increasingly important role in scientific inquiry, to the extent that computational science is now recognized as a third branch of physics, along with the pre-existing theoretical and experimental branches \citep{landau2005guide}.  Indeed, as theory becomes more complex and realistic, we often cannot directly solve theoretical models and derive predictions that should be measurable by experiment.  Computational science sits between theory and experiment, allowing us to understand the behavior and dynamics of complex theoretical models, and calculate predictions that can be used for experiment or hypothesis testing.

The problem of assessing simulation model quality is important enough that the Department of Energy and the Air Force Office of Scientific Research requested that the National Research Council study the foundations of verification, validation, and uncertainty quantification (VVUQ) activities for computational models in science and engineering.  Their draft report forms the basis of my approach to verification and uncertainty analysis in this research \citep{national2012Assessing}.

Verification answers the question, ``how accurately does a computational model solve the underlying equations of a theory for the observable quantities of interest.''  
Given that we know the true value of $\theta$ which drives our simulation runs, it is possible to calculate the expected number of variants at stationarity, and use this to verify that \tf is correctly implementing the WF-IA.   The expected number of traits is a good validation estimate because the number of variants present in a sample will be sensitive to the relative rates of copying and innovation events being handed correctly in the simulation code.  Errors in handling these events in software will be magnified across many individuals over many simulation steps.  

Since $\theta$ is known, the mean value of $K_n$ is well approximated by:

\begin{equation} 
\label{eq:expected-kn}
	\mathbb{E}_{\theta}(K_n) = \int _0^1\left(1-(1-x)^n\right)\frac{\theta }{x}(1-x){}^{\theta -1} dx
\end{equation}

Using Equation \eqref{eq:expected-kn}, I compared expected $K_n$ to the average of $k_n$  for a large sample of simulation runs.  To ensure that behavior is correct across a range of useful $\theta$ values, I performed multiple simulation runs at $\theta$ values ranging from 2 to 40, for 5000 generations in a simulated population of 2000 individuals.  Each parameter combination was represented by 3 simulation runs.  The initial transient behavior of the model is discarded from data analysis by skipping the first 750 generations, given the mixing time analysis by \citet{Watkins2010}.  At each time step in a simulation run, the simulator took a sample of 30 individuals and tabulated the traits held by those individuals, and recorded the value of $K_n$.  This yielded 408,478 samples across validation runs.  

Using Mathematica 8.0 with MathStatica 2.5 installed, I calculated expected values for each $\theta$ level used in simulation, employing Equation \eqref{eq:expected-kn}.  Table \ref{tab:validation-kn} compares the expected and observed values.  In all cases, the analytical results are extremely close to the observed mean $K_n$ values from simulation, and certainly well within 1 standard deviation.  Thus, I conclude that the \tf implementation of WF-IA employed in this study accurately represents the desired theoretical model.  

\subsection{Time-Averaging and Simulation Parameter Space}
\label{sec-ta-method}
Time-averaging was modeled in \tf by implementing a series of statistical ``windows'' within which trait counts were accumulated between time steps.  At the end of each temporal window, a sample of appropriate size is taken from the accumulation of trait occurrences, trait counts within that sample tabulated, and $K_n$ values recorded.  The simulator architecture allows an arbitrary number of temporal windows to be employed simultaneously (albeit with a small performance penalty for each window).  As a consequence, during a single simulation run, the simulator tracks both unaveraged statistics and the same statistics averaged over any number of ``assemblage durations.''  All trait samples taken in the simulator, whether unaveraged or for a specific assemblage duration, were also recorded to allow calculation of Slatkin's Exact test. Additionally, to facilitate analysis of time scales within the simulation model, for each trait the interval between entry and loss through drift was recorded.   In the simulation results reported here, trait samples were of uniform size 100.  Constant sample size removes the effect of different sample sizes on the reported results, although the interaction of the fixed sample size and the innovation rate will lead to cutoff behavior at very high $\theta$ values.  This is acceptable since the very highest $\theta$ values employed here are unrealistic for almost any prehistoric phenomena, and may be approached only for ``viral'' behavior on modern social networks.    

All simulations reported here were performed with a population size ($N$) of 2000 individuals, and simulation runs were conducted for the following values of $\theta$:  0.1, 0.25, 0.5, 1.0, 2.0, 5.0, and 10-100 at intervals of 10.  This range encompasses innovation rates that are very small, through populations in which a full 5\% of the population has a never\hyp{}before\hyp{}seen variant each generation.  Simulations were performed in several batches, with a core set of runs performed for 40,000 steps in order to determine the effects of long-duration \timeav, yielding simulated assemblages at a variety of windows ranging from 3 steps to 8000 steps (the exact durations sampled are given in the first column of Table \ref{tab:sample-size-kn}).  In order to increase the sample size of long-duration assemblages, a second set of simulation runs using the same parameters were done with only the five largest windows recorded (the short duration window sizes were discarded to avoid a flood of raw data beyond that needed for stable statistical analysis).  Finally, since the statistical behavior of the process at very small values of $\theta$ is highly variable, a third set of runs was performed to increase the number of samples for $\theta$ values between 0.1 and 1.0.  

Trait samples were post-processed outside the simulator environment, since calculation of Slatkin Exact tests within the simulator itself would slow down the primary simulation model by a large factor.   Montgomery Slatkin's original C language program was used in Monte Carlo mode to produce an estimate of $\mathbb{P}(E)$ for each sample of individuals.  I modified Slatkin's original \texttt{montecarlo.c} program to not require the data to be embedded in the source code, instead taking data as a command line parameter, and outputting only the $\mathbb{P}(E)$ value and $\theta$ estimate, to allow easy post-processing of the simulation output.\footnote{These modifications are available, along with all other analysis scripts, in the Github repositories http://github.com/mmadsen/saa2012, and the \tf source code.}  

The simulation results reported here, once post-processed, comprise 3,024,085 sample values for $K_n$, across the $\theta$ values listed above, and broken down across assemblage durations as in Table \ref{tab:sample-size-kn}, and 1,113,134 Slatkin Exact test results for the same combinations of $\theta$ and assemblage duration.

\section{Results}
\label{sec:results}

\begin{figure*}
	\includegraphics[angle=90, scale=0.75]{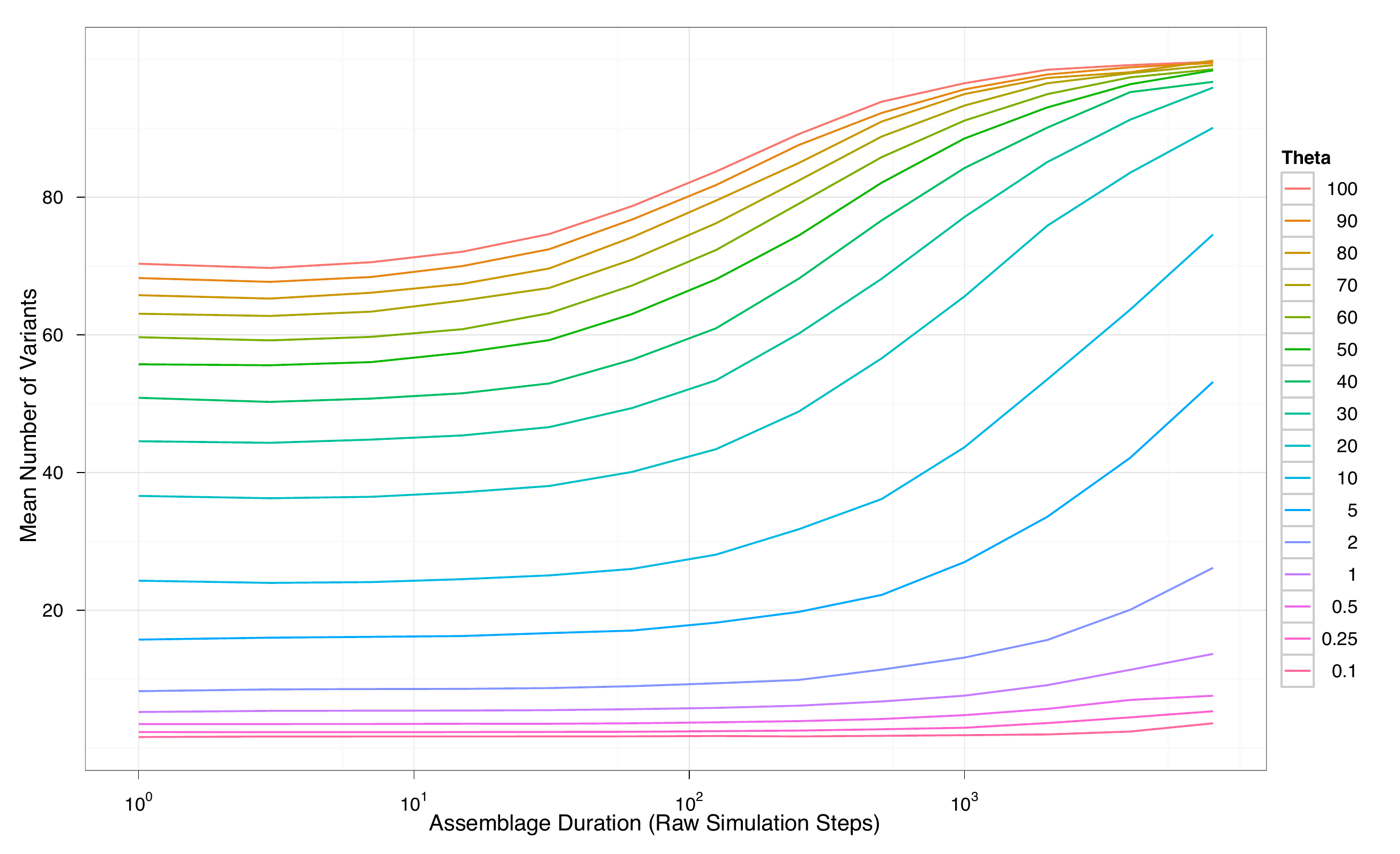}
	\caption{Mean value of $K_n$ for \timeavd samples, plotted against assemblage duration in simulation steps, for each level of $\theta$ in the study.  Note that the ``onset'' of \timeav effects (as measured by increased $K_n$), is quite gradual at low $\theta$, while high innovation rates display increased richness with very minor amounts of \timeav.}
	\label{fig:unscaled_kn}
\end{figure*}

Simple inspection of the relationship between assemblage ``duration'' (i.e., accumulation interval) and the average number of variants ($K_n$) in a sample of size 100, shows a strong \timeav effect (Figure \ref{fig:unscaled_kn}).\footnote{Here, the time axis represents raw simulation steps, each of which represents $N = 2000$ copying events within the population.  This is the only figure in this paper which uses raw simulation time steps as the time variable.}  Temporal aggregation of the results of transmission inflates the number of variants we see in a sample, with greater effect as the population innovation rate ($\theta$) increases.  The effect is very small at low theta values (i.e., when the process is drift-dominated, $\theta < 1.0$) and requires long accumulation of copying events to have a measurable effect upon mean $K_n$.  
Conversely, inflated $K_n$ appears at fairly short duration as theta increases.  

Simulation steps (or ``generations'') represent an arbitrary time scale with respect to the chronological time archaeologists can (with effort) measure.   In order to understand the effects of \timeav on archaeologically-relevant time scales, it will be useful to rescale simulation time by some factor which is observable as a function of artifact class duration in the depositional record.  I take up this issue further in Section \ref{sec:conclusions}, but the ideal time scale would be the mean duration of artifact classes in the classification system being used in a given empirical study.   I do not explicitly model archaeological classification in the present results, but a related measure is the lifespan of the traits being transmitted within the simulated population.  

\subsection{Time Scales and \Timeav}

The ``mean trait lifetime'' in WF-IA is a direct consequence of the balance between innovation and loss of traits to drift, in a fixed-size population.  At the quasi-stationary state, the population will fluctuate around a mean number of traits, as individual traits enter and leave the population constantly.  This implies that at stationarity, if we add traits at a higher rate due to migration or innovation, more traits must be lost to drift each generation.  WF-IA thus satisfies a balance equation characterizing the average number of variants ($\bar{n}$)\citep{ewens1964maintenance}:

\begin{equation}
\label{eq:nug}
\frac{\bar{n}}{\bar{t}} = \theta
\end{equation}
where $\bar{t}$ represents the average number of generations that a new trait lasts in the population before its loss to drift (i.e., the mean trait lifetime).  

An exact expression for mean trait lifetime has not been derived from the transition probabilities of the WF-IA Markov chain \citep{ewens1964maintenance}, but it can be approximated by summing the average amount of time that a trait within a population spends at specific frequencies (i.e., mean sojourn times).  \citet[Eq. 3.20]{Ewens2004} gives the following approximation:
\begin{equation}
	\label{eq:mean-trait-lifetime}
	\bar{t} \approx \mathbb{E}(t_i) = \sum _{j=1}^{\infty } \frac{2N}{j(j-1+ \theta )}(1-(1-p)^j)
\end{equation}

Since $\theta$ is in the denominator of the summation, increasing the population rate of innovation reduces the mean trait lifetime by decreasing the amount of time any specific trait spends at a given frequency, and thus the total amount of time a trait spends in the population before being lost to drift.   

Table \ref{tab:mean-trait-lifetime} lists the observed mean lifetime of traits for each level of $\theta$ employed in this study, and the expected value as calculated using Equation \ref{eq:mean-trait-lifetime}.  The observed values are systematically lower than the expected values, which reflects slightly faster loss of traits due to drift in a finite and small population compared to the large populations often studied in population genetics \citep{ewens1964maintenance,kimura1964number}.  Examination of Figure \ref{fig:unscaled_kn} appears to show that the onset of \timeav effects, however small, occurs around the time scale of the mean trait lifetime, for values of $\theta \geq 1.0$.  This outcome is sensible given the enhanced probability of longer duration samples incorporating new variants in the sample due to innovation.  In the analyses to follow, I scale the time variable by the mean trait lifetime, displaying assemblage duration as a multiple of this value.  Thus, for the remainder of this paper, a scaled assemblage duration of 100 will indicate 100 times the mean trait lifetime at that specific $\theta$ value.  For example, if we are examining results at $\theta = 5.0$, a scaled duration of 100 would indicate $12.43 * 100 = 1243$ simulation steps.  

\subsection{Neutrality Testing}
\label{sec:slatkin-ta-effects}

The Slatkin Exact test for neutrality, discussed in Section \ref{sec:neutrality-test}, determines the ``tail'' probability for a sample of size $n$, with observed number of traits $k$, to be derived from the Ewens Sampling Formula (Equation \ref{eq:conditional-esd}).  The test employed in this study is Slatkin's Monte Carlo version, which allows the use of larger sample sizes, using random selection to create unlabeled configurations from the ESD to compare against the observed values.  The resulting tail probability is converted into a standard hypothesis test by selecting an $\alpha$ value.  For purposes of this study, I considered the upper and lower 5\% of tail probabilities to indicate that a sample was probably not derived from a neutral transmission model, leading to $\alpha = 0.10$.  

Given this $\alpha$ level, we should expect roughly 5\% of the samples taken from a pure neutral copying process to fall into each of the the upper and lower tail regions, and thus for a Slatkin Exact test to reject the null hypothesis of neutrality.  Roughly 90\% of the samples we take from the neutral WF-IA process should fall between $0.05 < p < 0.95$ and thus lead to acceptance of the null hypothesis.  This experimental setup also implies the limited utility of performing a single neutrality test on a single sample of class counts or frequencies, as has been archaeological practice by necessity.  A single Slatkin exact test with $\mathbb{P}_e$ value of, say, 0.96, would constitute some, but relatively weak, evidence of non-neutrality.  Better practice would be taking many samples from a large assemblage or multiple collections and calculating independent Slatkin tests for each sample, and examining their distribution.  

If \timeav has no effect on the validity of the Slatkin Exact test employed against temporally aggregated samples, we would expect the fraction of samples in the two tails (upper and lower 5\% in this case) to equal 10\%.  Anything over 10\% would constitute evidence of extra Type I errors, since we know the samples to have been generated by a process meeting the definition of the null hypothesis.  Therefore, after post-processing the simulation output to produce Slatkin tests as described in Section \ref{sec-ta-method}, I tabulated the fraction of Slatkin Exact tail probabilities that exceeded the expected 10\% tail population.  These are, in other words, ``excess'' failures of the Slatkin Exact test, beyond those expected by the probability distribution itself.  For each $\theta$ level, and for each \timeav duration, the mean ``excess'' failure rate was computed, from the 1,113,134 raw Slatkin Exact test results generated in the simulation study.  

\begin{figure*}
	\includegraphics[angle=90, scale=0.75]{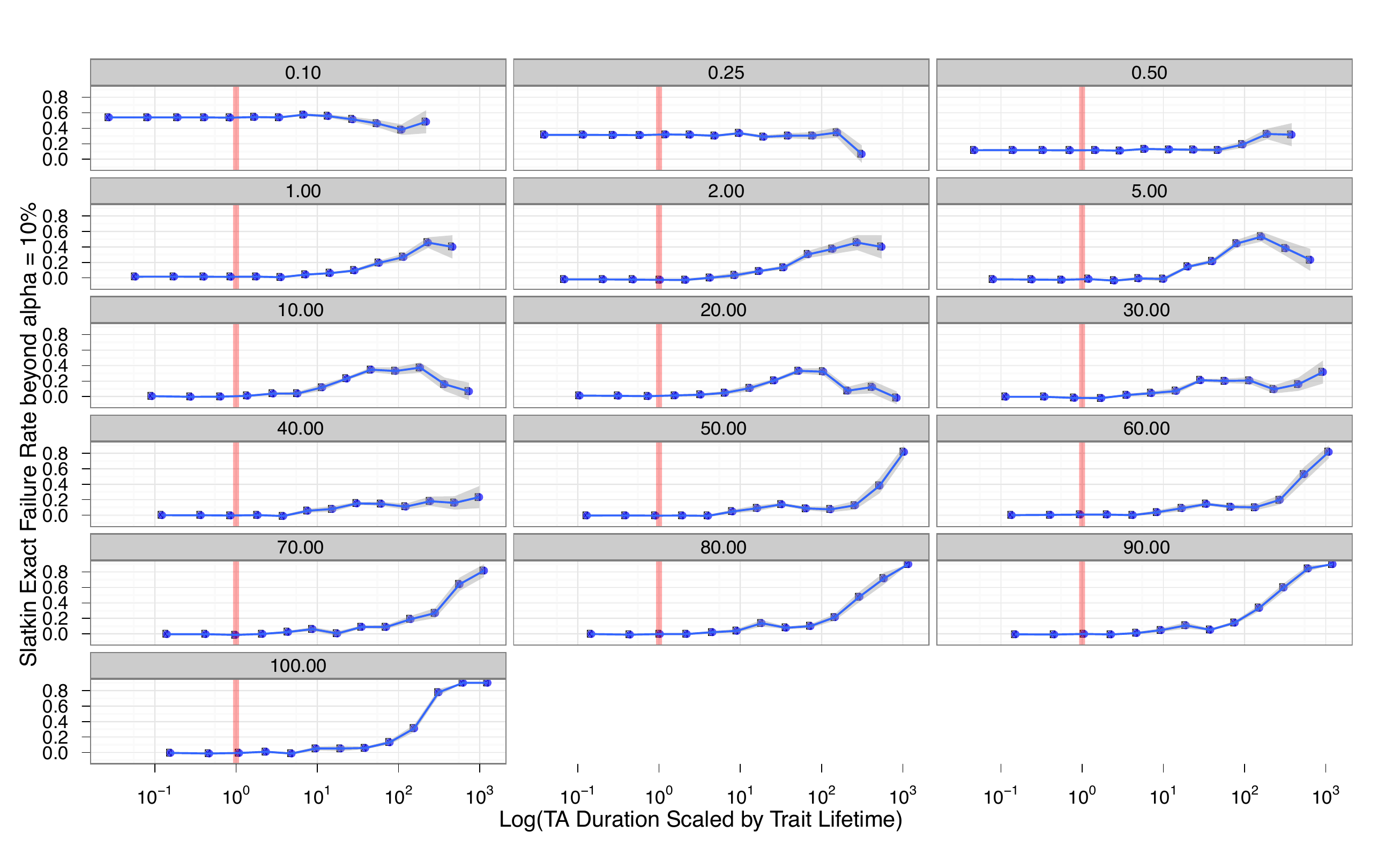}
	\caption{Slatkin Exact test failure rate (above the expected 10\% given two-tailed test with $\alpha = 0.10$, plotted against \timeav duration scaled by mean trait lifetime, for each level of $\theta$ in the simulation study. The red vertical line indicates the mean trait lifetime for that $\theta$ value, and the shaded region encompasses the standard error of the estimates for mean failure rates at each duration.}
	\label{fig:extra-slatkin-failures-by-scaled-duration}
\end{figure*}

Figure \ref{fig:extra-slatkin-failures-by-scaled-duration} depicts the relationship between the excess failure rate, and \timeav duration scaled by the mean trait lifetime (as previously described).  The mean trait lifetime is indicated by a vertical red bar in each graph.  Three major results are apparent.  First, at values of $\theta \geq 1.0$, the excess failure rate in non-time-averaged data is zero, as one would expect, and then begins to increase (albeit slowly) as the \timeav duration of samples exceeds the mean trait lifetime.  In some cases, such as $\theta = 5.0$, the Slatkin Exact test continues to be accurate given the chosen $\alpha$ value through samples which are aggregated for 10 times the mean trait lifetime.  But in all cases, with sufficient \timeav, the Slatkin Exact test begins to suffer from increased Type I error, reporting an ever increasing fraction of samples as not derived from a neutral transmission process.  The extreme situation is seen at very high rates of innovation, where nearly every test fails, at high levels of \timeav.  These failures are caused by saturation of a finite sample with singleton traits, causing the sample to display too much evenness in frequency to have been the result of sampling from the Ewens Sampling Formula.  But unrelated to this saturation effect, there is considerable failure in employing the Slatkin Exact test to detect neutrality.  For example, at $\theta = 5.0$, at 100 times the mean trait lifetime, approximately 70\% of all samples appear in the tail region of the distribution, compared to the expected 10\%.  Clearly, the Slatkin Exact test is not robust for long-duration assemblages.    

Second, at low $\theta$ values, the test results show excessive Type I error, even without \timeav.  There are several potential causes.  It is possible that the WF-IA process had not reached quasi-stationarity by 750 time steps, when sampling began.  This would mean that the effects of initial trait assignment might still be present and skewing the frequency distribution of traits.  Second, the Slatkin test is sensitive to the number of rare or singleton traits given the sample size, and in a small population (2000 individuals) with a low innovation rate (e.g., $\theta = 0.1$), counts of rare traits could be unstable.  This would not typically be the case in samples from large populations or entire species.  I do not consider the cause of this anomaly further in this paper, but it warrants further simulation study.   

In general, with long-duration assemblages, archaeologists should be careful interpreting the results of neutrality tests adopted from population genetics.  The effect seen here can be summarized as:  with significant \timeav, trait frequencies generated by unbiased \ct can falsely appear to be non-neutral and thus driven by bias or selection (Type 1 error).  The longer the duration of an assemblage with respect to the mean trait lifetime, the larger the probability of a Type 1 error.  With sufficient duration, in fact, the probability of a Type 1 error becomes virtually certain, and the Slatkin Exact test loses any discriminatory power.  In summary, if one were to employ Slatkin's test to examine the hypothesis of neutrality in long-duration archaeological deposits, one would overwhelmingly come away with the impression that most \ct was biased, either towards conformity or a pro-novelty bias -- regardless of the underlying process occurring during prehistory.

\subsection{Theta Estimation and Innovation Rates}
\label{sec:theta-estimation-results}

There would be considerable value in estimating the population-level innovation rate ($\theta$) from sample data if it could be done accurately.  As discussed in Section \ref{sec:theta-estimation-theory} above, such estimates are usually biased and have large variance.  In this section, I examine the effects of \timeav upon theta estimates generated from the samples taken to perform neutrality tests in the previous section.  For each of the 1.1 million samples of variants (distributed across actual theta values and assemblage durations), I calculated theta estimates given Watterson's approximation \citep{durrett2008}:

\begin{equation}
\label{eq:watterson-theta-est}
	\hat{\theta} \approx \frac{k_n}{log\;n}
\end{equation}

For each combination of actual theta and assemblage duration, theta estimates were averaged, to give a mean estimated theta value ($\mathbb{E}(\hat{\theta})$), and its standard deviation.  The results are shown in Table \ref{tab:estimated-theta-unaveraged}.   There are two regions of behavior apparent in the table, corresponding to drift- versus innovation-dominated dynamics.  At and below $\theta = 1.0$, estimated values are higher than the actual $\theta$ used to generate samples, and above 1.0, theta estimates begin to systematically lag below the actual theta value.  Overestimation at $\theta \leq 1.0$ matches the simulation results by \citet{ewens1974note}, although the authors did not simulate innovation rates above 2.0 (a large value in most genetic situations).  In addition to being biased, theta estimation appears to be even \emph{approximately} accurate only within a narrow range of values around $\theta = 1.0$.  

\begin{figure*}
	\includegraphics[scale=0.85]{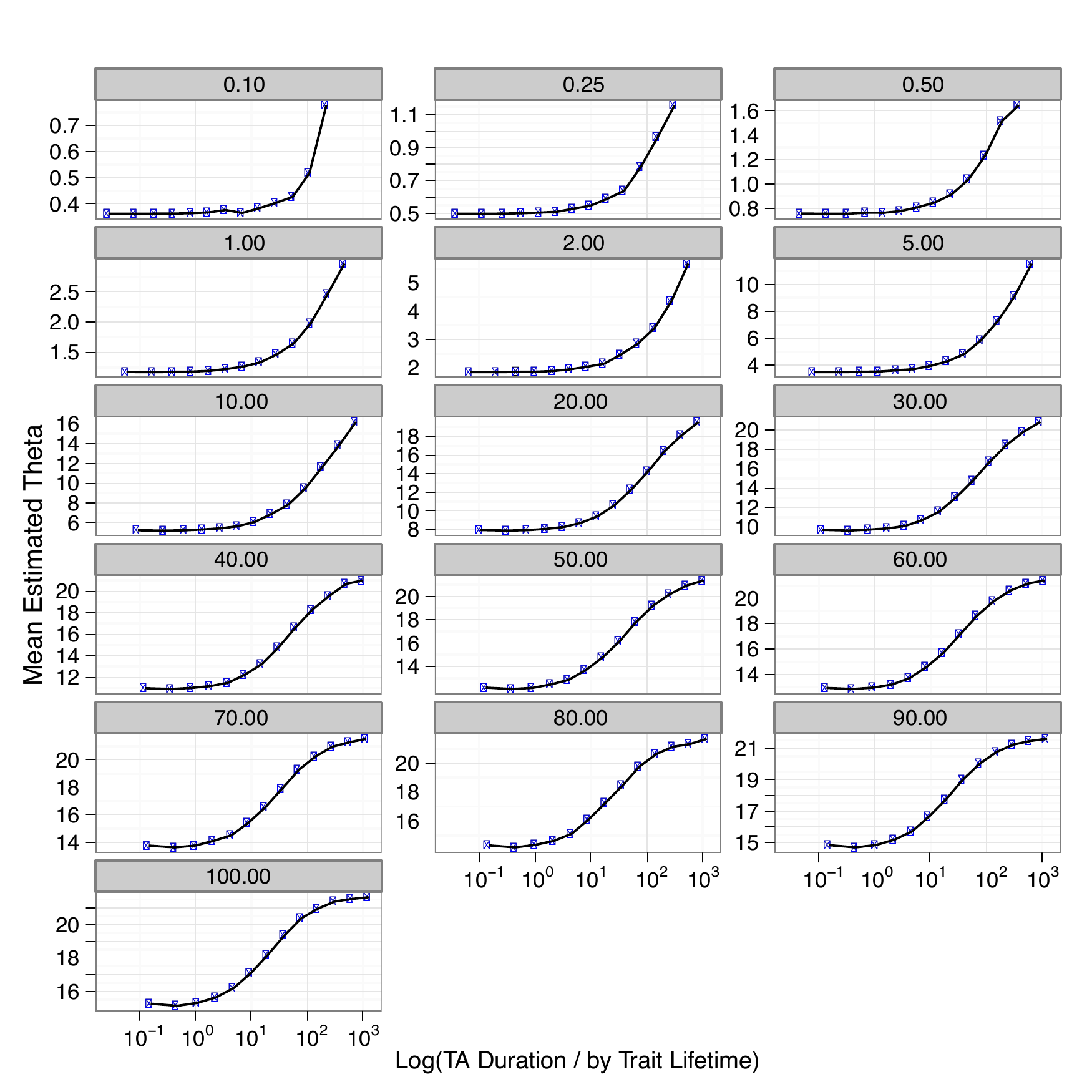}
	\caption{Estimates of mean population innovation rate ($\mathbb{E}(\hat{\theta})$) from samples ($n = 100$) taken for neutrality tests, using the approximation by \citet{watterson1975number}.  Plotted against assemblage duration, for each level of actual innovation rate used in simulation runs. }
	\label{fig:theta-estimates}
\end{figure*}

Figure \ref{fig:theta-estimates} examines estimates of theta by \timeav duration scaled by the mean trait lifetime, for each level of actual $\theta$ used in the simulation runs.  The pattern evident in synchronic or unaveraged samples carries over to \timeavd assemblages:  below $\theta \leq 1.0$, theta estimates are larger than the actual values, and increase in a non-linear fashion with assemblage duration.  Above 1.0 but below about 30.0, theta estimates begin below the actual value, cross the actual value, and continue to accumulate as assemblage duration increases.  Finally, at the very highest innovation rates, in a sample size 100, theta estimates are always drastic underestimates of the actual value, even with long assemblage duration increasing the accumulation of traits.  

\begin{figure*}
	\includegraphics[scale=0.85]{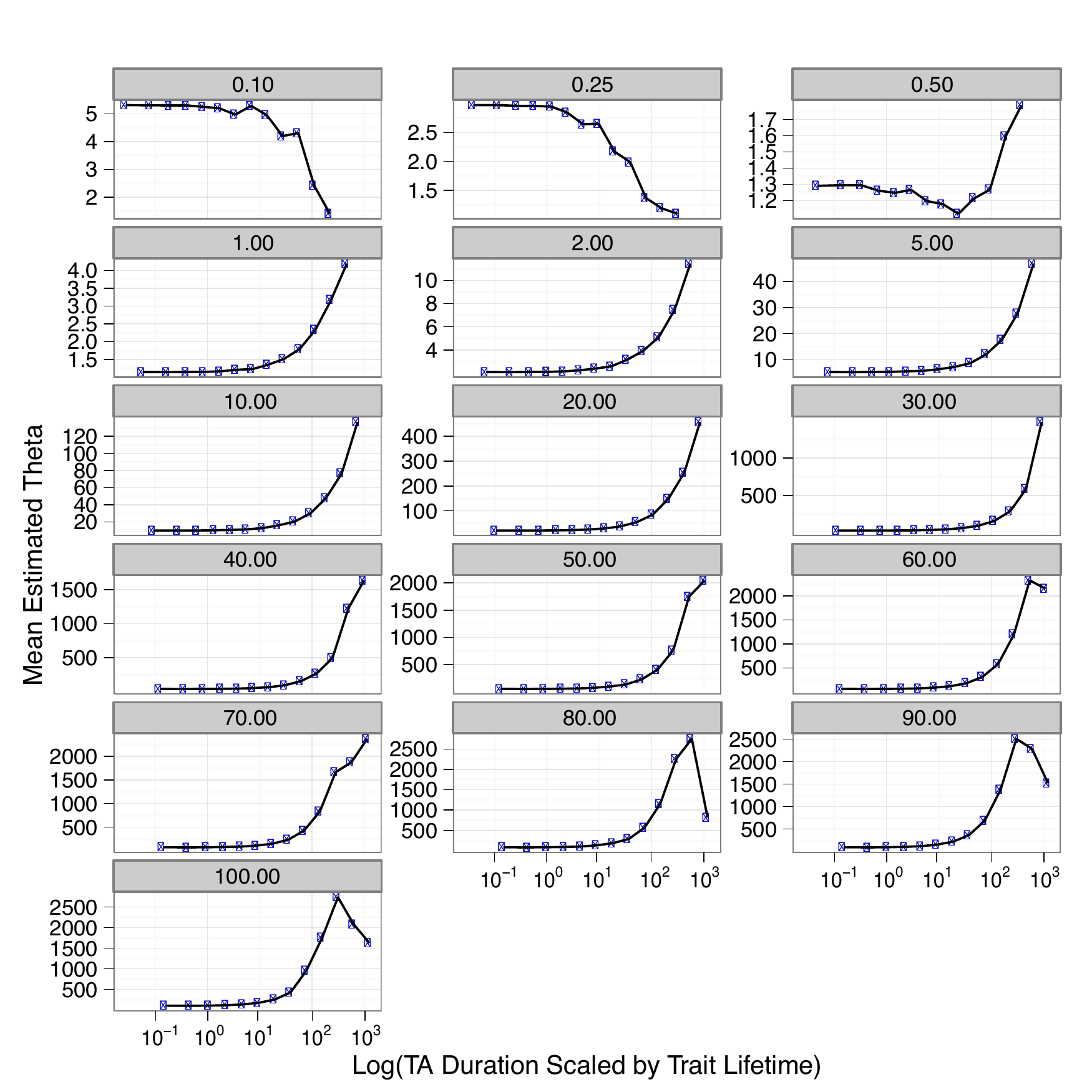}
	\caption{Estimates of mean population innovation rate ($\mathbb{E}(\hat{\theta})$) from samples ($n = 100$) taken for neutrality tests, using results from Montgomery Slatkin's neutrality test software.  Plotted against assemblage duration, for each level of actual innovation rate used in simulation runs. }
	\label{fig:theta-estimates-slatkin}
\end{figure*}

The Slatkin Exact test software also provides an estimate of $\theta$, finding the maximum likelihood value of theta when $K_n$ is set in Equation \ref{eq:expected-kn} to equal the observed value (this is the $t_e$ statistic introduced to archaeological usage in \citealp{Neiman1995}).  Figure \ref{fig:theta-estimates-slatkin} depicts the Slatkin theta estimates by \timeav duration scaled by the mean trait lifetime, for each level of actual $\theta$ used in the simulation runs.  One interesting difference between Figure \ref{fig:theta-estimates} and the Slatkin theta estimates is that the latter are more accurate for actual $\theta \geq 1.0$ than the Watterson approximation, in unaveraged assemblages.  Unfortunately, with increased assemblage duration, estimates explode to much larger values than those calculated by the Watterson approximation (i.e., $\theta \approx 1500$ for true $\theta = 30$ at maximum assemblage duration of 1000 times the mean trait lifetime, compared to the \emph{underestimate} of approximately 22 in Figure \ref{fig:theta-estimates}). 

In short, estimation of population-level innovation rates from samples of artifacts using either estimation method are inaccurate, and the \timeav effect of accretional deposition renders such estimates even more inaccurate.  Clearly, such values cannot be used as actual indications of innovation rate or to ``work backward'' towards past population sizes.  And without fairly precise control over assemblage duration, the use of $t_e$ as a relative diversity measure between assemblages (in the manner common to archaeological applications) is highly suspect.  In the next section, I turn to $t_f$, the other common diversity measure in archaeological studies, which does not require an estimate of $\theta$, employing instead the variant frequencies directly.    

\subsection{Diversity Measures}

Much of the current effort in distinguishing biased and unbiased transmission models rely upon trait evenness and the shape of frequency distributions, given Alex Bentley's application of power-law distributions to both ancient and contemporary data sets \citep{Bentley2003,bentley2004random,Bentley2007,bentley2007fashion, bentley2009physical,8913,8914}.  One of the ways that unbiased and ``conformist'' models of \ct differ is in the expected amount of variation.  Compared to unbiased transmission, conformism of even a mild degree tends to strongly concentrate adoption onto a very small number of traits \citep{Mesoudi2009}.\footnote{This is especially the case when conformist transmission is implemented in simulations as a ``global'' rule where only the most common trait is copied during ``conformist'' copying events, rather than weighting all traits by their relative popularity.  Very little work has been done to compare the results from different methods of simulating biased transmission models.  This is a topic which would benefit greatly from additional research.}  It is difficult, however, to interpret the absolute number of traits ($K_n$) without knowledge of the population size, so \citet{8977} employed diversity measures instead in his classic examination of conformist transmission in Southwest pottery.  

The most commonly used measure in the archaeological literature on \ct is $t_f$ (Equation \ref{eq:tf-formula}), since it is related to Wright's original measures of heterozygosity and thus associated directly with the historical development of the Wright-Fisher model.  But it is useful to normalize the results of $t_f$ between 0 and 1 so that we can compare different levels of theta and assemblage durations easily, in the same way that statisticians occasionally employ coefficients of variation or normalize covariances into correlation coefficients.  Equation \ref{eq:iqv-formula} does exactly this, and is called the ``index of qualitative variation'' (IQV) \citep{wilcox1973indices}.

\begin{figure*}
	\includegraphics[angle=90, scale=0.75]{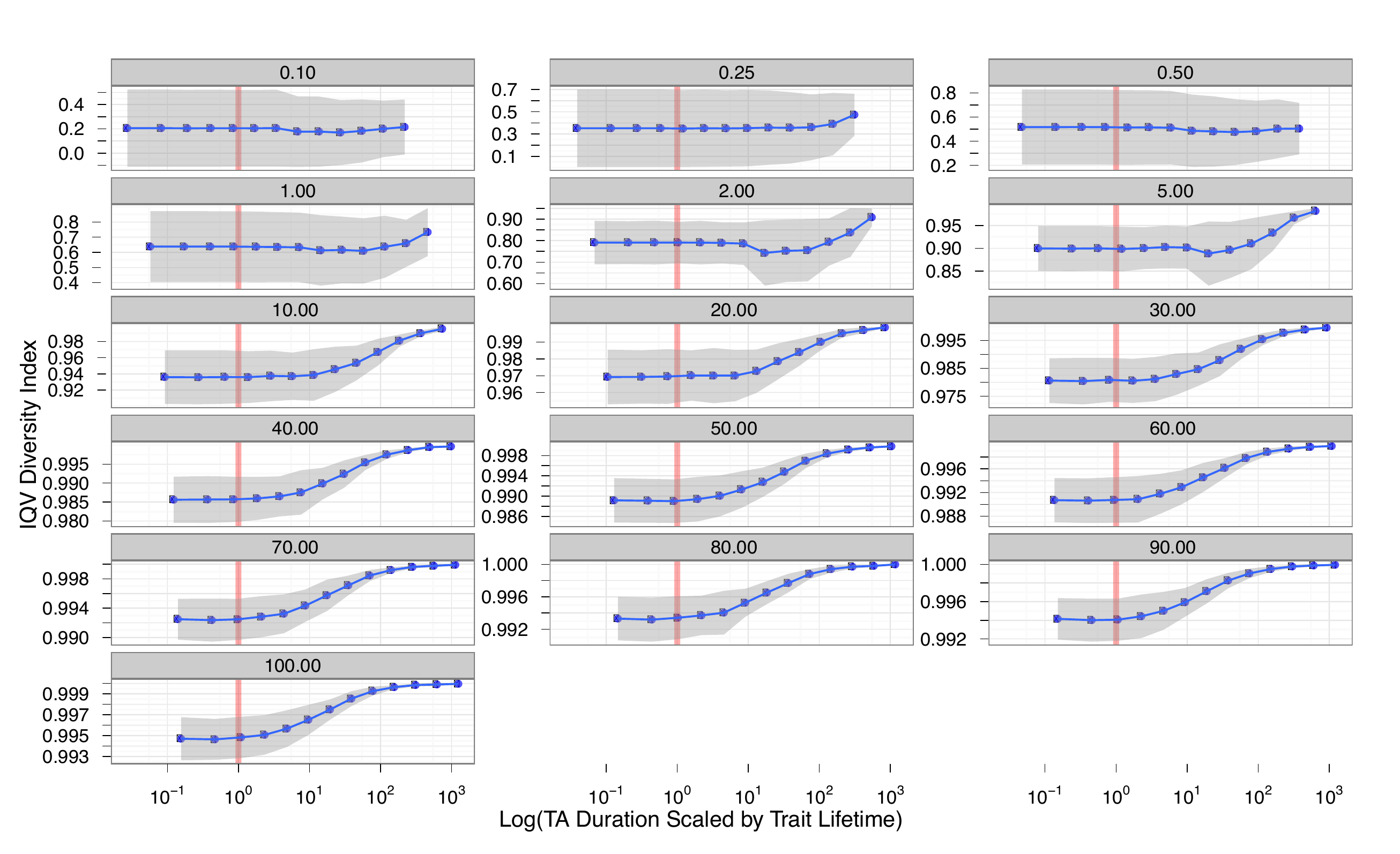}
	\caption{IQV diversity index, derived from samples of size 100, plotted against \timeav duration scaled by mean trait lifetime, for each level of $\theta$ in the simulation study. The red vertical line indicates the mean trait lifetime for that $\theta$ value.}
	\label{fig:iqv-est-by-scaled-duration}
\end{figure*}

Figure \ref{fig:iqv-est-by-scaled-duration} displays the relationship between the IQV for samples of size 100, and \timeav duration scaled by mean trait lifetime, as before.  IQV values range from 0.0, if only a single trait occurred within a sample (which happens in simulations with very low innovation rates), through 1.0, which indicates that traits are perfectly evenly distributed within a sample.  Even at the highest innovation rate studied, values of 1.0 were not seen in \emph{unaveraged} samples from the simulation runs.  It is apparent that \timeav can yield greater evenness among trait frequencies, although the plateau in IQV values  seen at high $\theta$ and high assemblage duration is a function of the saturation of $K_n$ in a finite sample seen above.  At very low innovation rates ($\theta \ll 1.0$), \timeav in contrast seems to have little effect on the dispersion of trait frequencies, with one or a very few traits always dominating a sample.  

In between, when innovation rates are sufficient to guarantee at least one innovation on average per model generation ($\theta = 1.0$) but fewer than 10, there is non-monotonic behavior apparent in the IQV index.  For example, at $\theta = 2.0$, \timeav has no effect on IQV until duration is 10 times the mean trait lifetime (\mtl), at which point assemblages begin to appear \emph{less even} in frequency distribution, until about 100 times the mean trait lifetime, when evenness begins to steadily increase.   This effect is interesting, since it suggests that we cannot easily compare diversity indices between assemblages unless we control for duration or have independent evidence concerning innovation rates.

\section{Discussion and Conclusions}
\label{sec:conclusions}

When we examine the effects of \timeav on the sample properties of unbiased transmission, using the mean lifetime of traits as our fundamental time scale, several lessons for practical applications emerge.  First, it appears that assemblages with very small amounts of temporal aggregation display little of the distributional alterations that characterize long-duration assemblages.  Statistical tests of neutrality and diversity measures, and thus arguments based on them, can probably be used with care.   Second, estimates of population-level innovation rate derived from Ewens's sampling formula are biased (and therefore inaccurate), and become seriously inaccurate with increased assemblage duration.  Archaeologists should strongly reconsider using $t_e$ or other theta estimates even in relative comparisons, and should definitely not consider such estimates to reflect the innovation rate or population size present in the prehistoric population.  Third, for assemblages that have a duration longer than the mean trait lifetime, it is important to measure and control for the relative duration of assemblages when comparing statistical results across samples.  Without doing so, we cannot interpret relative differences of diversity indices or trait richness values as indicative of different modes of transmission.  

One caveat to the above is that such effects refer specifically to \emph{assemblage} level data, composed of many artifacts deposited over time.  Artifact-scale analysis, where the attributes under analysis come together in a short period of time, and where single artifacts comprise the counting unit for transmission studies, will not necessarily suffer the quantitative effects described here, or would suffer no measurable \timeav effects if the assemblage durations were short compared to the lifetime of traits.  A good example of this is Jonathan Scholnick's chapter in the present issue, expanding on his previous research into \ct in Colonial America through gravestones \citep{premo2011spatial,scholnick2010apprenticeship}, where his samples cover 10 year periods based on the death dates carved on each stone.  

Furthermore, while the mean lifetime of transmitted information plays a central role in establishing a ``natural'' time scale over which \timeav affects  unbiased transmission, this time scale is not an archaeological one.  This discrepancy in time scales arises because the abstract ``traits'' of our models are not equivalent to the classification units employed by archaeologists.  This is not a trivial difference, and is one that is rarely even discussed in archaeological applications of \ct models.  Instead, we frequently act as if  ``traits equal types,'' despite occasional acknowledgement of the difference. 

But we have no direct empirical access to the information prehistoric populations were learning, teaching, and imitating.  We will never find ``units of transmission'' in any empirical sense for archaeological applications of \ct models, and we have no warrant to equate our models of prehistoric information flow with the classes we use to observe it today.  Long ago, \citet{8874} recognized that when anthropologists study the ideas held within a social group under study, what is actually being studied are the ideas we \emph{construct} about the ideas individuals in other cultures may have had.  \citet{Dunnell1971} systematized this distinction, pointing out that we always operate with analytic classes whose construction is done by archaeologists, for archaeological purposes.  These classes serve as a ``filter'' by which we detect patterns in artifact assemblages, which reflect patterns in the information which flowed within past populations.  There is no ``natural'' set of classes to employ in studying \ct, but we often forget to incorporate this fact into our analyses.  Linking the time scale over which variation entered and left a prehistoric population, and the time scale over which archaeological classes appear and then exit the archaeological record will involve further research on the relationship between transmission dynamics and complex, multi-dimensional archaeological classes.  Such research is essential to connect the abstract quantities described by theoretical models, to observable aspects of the archaeological record.  

These results paint a fairly gloomy picture of almost all of the standard variables archaeologists have used since Neiman's \citeyearpar{Neiman1995} pioneering work.  One wonders why empirical studies using diversity measures, innovation rate estimates, or neutrality tests appear to ``work'' and give sensible results?  One possibility, of course, is that some studies don't yield the expected results.  We see this, possibly, in a fascinating analysis by \citet{steele2010ceramic}.  The authors employed ceramic classes that appeared to be non-neutral and subject to selection or biased transmission.  Yet Slatkin exact tests were unable to rule out the null hypothesis of neutrality.  I do not present an analysis of conformist transmission under \timeav in this article, but using \tf I see evidence that temporal aggregation has the opposite effect on Slatkin exact tests in populations with weak conformist biases:  neutrality tests suffer increased Type II error, making it more likely that we will accept a null hypothesis of neutrality when the opposite is the case.  

Another possibility is that certain variables may retain their distributional character, but have their values inflated by temporal aggregation.  In such situations, there would be no reason to reject the neutral model, but inferences about the values of parameters would be inaccurate.  Even if investigators did not rely upon the absolute value of parameters, frequently such inferences (e.g., diversity values) are employed as relative comparisons between assemblages.   I suspect that this has occurred in a number of published studies, but few \ct applications include detailed information concerning assemblage duration, so it is difficult to redo the researcher's original hypothesis tests with temporal controls, without going back to original field information or reports.  Clearly, both possibilities may also occur in some situations.  

As archaeological usage of \ct theory becomes more frequent and we move from proof-of-concept studies to demanding interpretive accuracy from our models, methodological research is essential to ensure that our applications are empirically and dynamically sufficient.  The present study focused on a necessary first step in such method development, developing an understanding of the effect of \timeav in accretional assemblages upon the observable variables in neutral \ct models.  The results demonstrate that frequently employed statistics, such as $t_e$, are highly inaccurate and biased when measured in \timeavd assemblages, and that neutrality tests are subject to enough additional Type I or Type II error that the results can be systematically misleading.  Clearly, in order to apply \ct models to diachronic data derived from \timeavd assemblages, we need to develop observational tools and methods suited specifically to the archaeological record, instead of simply borrowing statistical methods and models from theoretical population biology.

\section{Acknowledgements}
This paper was originally presented in a symposium titled \emph{``Recent Developments in Cultural Transmission Theory and its Applications''} at the 2012 Annual Meeting of the Society for American Archaeology, Memphis, TN.  The author wishes to thank Kristen Safi, the organizer, for the opportunity to participate, and Carl P. Lipo, Fraser Neiman, James Feathers, Jonathan Scholnick, and Michael J. O'Brien for comments on drafts of this paper.

\clearpage

%

\clearpage

\begin{table}[ht]
\begin{tabular}{|c|c|c|}
\hline
TA Duration & Min Sample Size & Max Sample Size \\ 
  \hline
  1 & 130494 & 247491 \\ 
    3 & 4497 & 43494 \\ 
    7 & 1926 & 18639 \\ 
   15 & 897 & 8694 \\ 
   31 & 435 & 4209 \\ 
   62 & 216 & 2103 \\ 
  125 & 105 & 1038 \\ 
  250 & 516 & 981 \\ 
  500 & 255 & 486 \\ 
  1000 & 114 & 228 \\ 
  2000 &  57 & 114 \\ 
  4000 &  27 &  54 \\ 
  8000 &  12 &  16 \\ 
  \hline
\end{tabular}
\caption{Breakdown of sample sizes for analysis of trait richness ($K_n$), by size of time-averaging ``window.''  Some values of $\theta$ required larger numbers of simulation runs to achieve stable result, thus the difference between samples sizes at the same TA duration.}
\label{tab:sample-size-kn}
\end{table}

\clearpage

\begin{table}[ht]
	\begin{tabular}{|c|c|c|c|}
		\hline
		Theta & $\mathbb{E}(K_n)$ & Simulated $\bar{K}_n$ & Sim. Stdev $K_n$ \\
		\hline
		2 & 6.054 & 6.511 & 1.838 \\
		4 & 9.022 & 8.991 & 2.269 \\
		8 & 12.869 & 12.616 & 2.464 \\
		12 & 15.397 & 15.306 & 2.571 \\
		16 & 17.228 & 17.187 & 2.569 \\
		20 & 18.629 & 18.737 & 2.486 \\
		40 & 22.601 & 22.693 & 2.253 \\
		\hline
	\end{tabular}
	\caption{\label{tab:validation-kn}Comparison of expected $K_{n}$ from \eqref{eq:expected-kn} with simulated values from WF-IA model, for $\theta$ values from 2 to 40.  Total sample size across $\theta$ values is 408,478 samples of size 30.  }
\end{table}

\clearpage

\begin{table}[ht]
\begin{tabular}{|c|c|c|}
  \hline
Theta & Mean Trait Lifetime & $\mathbb{E}(t_i)$ \\
  \hline
0.10 & 36.54 & 36.89 \\ 
  0.25 & 25.61 & 24.05 \\ 
  0.50 & 21.10 & 19.97 \\ 
  1.00 & 17.31 & 17.21 \\ 
  2.00 & 14.57 & 15.21 \\ 
  5.00 & 12.43 & 13.05 \\ 
  10.00 & 10.83 & 11.57\\ 
  20.00 & 9.50 & 10.16 \\ 
  30.00 & 8.68 & 9.36 \\ 
  40.00 & 8.12 & 8.79 \\ 
  50.00 & 7.72 & 8.36 \\ 
  60.00 & 7.36 & 8.01\\ 
  70.00 & 7.08 & 7.72 \\ 
  80.00 & 6.83 & 7.46 \\ 
  90.00 & 6.60 & 7.42 \\ 
  100.00 & 6.40 & 7.05 \\ 
   \hline
\end{tabular}
\caption{Mean lifetime (in model generations) of traits, by $\theta$, along with analytical approximation from Equation \ref{eq:mean-trait-lifetime}.}
\label{tab:mean-trait-lifetime}
\end{table}

\clearpage

\begin{table}[ht]
\begin{tabular}{|c|c|c|}
  \hline
$\theta_0$ & $\mathbb{E}(\hat{\theta})$ & $\sigma(\hat{\theta})$ \\ 
  \hline
0.10 & 0.36 & 0.21 \\ 
  0.25 & 0.50 & 0.26 \\ 
  0.50 & 0.76 & 0.33 \\ 
  1.00 & 1.17 & 0.42 \\ 
  2.00 & 1.85 & 0.51 \\ 
  5.00 & 3.49 & 0.67 \\ 
  10.00 & 5.23 & 0.87 \\ 
  20.00 & 7.93 & 0.95 \\ 
  30.00 & 9.70 & 0.99 \\ 
  40.00 & 10.99 & 0.99 \\ 
  50.00 & 12.19 & 1.00 \\ 
  60.00 & 12.94 & 1.01 \\ 
  70.00 & 13.76 & 0.97 \\ 
  80.00 & 14.32 & 0.98 \\ 
  90.00 & 14.85 & 0.94 \\ 
  100.00 & 15.27 & 0.95 \\ 
   \hline
\end{tabular}
\caption{Mean Estimated Theta ($\mathbb{E}(\hat{\theta})$) from Samples (n=100) compared to actual values employed in simulation models ($\theta_0$), without any time-averaging.}
\label{tab:estimated-theta-unaveraged}
\end{table}

%
%


\end{document}